\documentclass[useAMS,usenatbib,a4]{mn2e}
\usepackage{amsmath}
\usepackage{amssymb}
\usepackage{amsbsy}
\usepackage{graphicx}

\def\kpc{\,{\rm kpc}}
\def\pc{\,{\rm pc}}

\def\kms{\,{\rm km\,s^{-1}}}

\def\Gyr{\,{\rm Gyr}}

\def\percent{\text{ per cent}}

\begin{document}
\title[Velocity dispersion determination]{Determining the velocity dispersion of the thick disc}
\author[J. Sanders]{Jason Sanders\thanks{E-mail: jason.sanders@physics.ox.ac.uk}\\
Rudolf Peierls Centre for Theoretical Physics, Keble Road, Oxford OX1 3NP, UK}

\pagerange{\pageref{firstpage}--\pageref{lastpage}} \pubyear{2012}
\label{firstpage}

\maketitle

\begin{abstract}
We attempt to recover the mean vertical velocity and vertical velocity dispersion as a function of the Galactic height for a sample drawn from a realistic Galaxy distribution function by following the method presented in \cite{MB2012}. We find that, for the sample size used, the observational error in the velocities is much smaller than the Poisson noise which has not been accounted for by \citeauthor{MB2012} We repeat the analysis on a large number of samples to estimate the contribution of the Poisson noise and to uncover any systematics. We find that the dispersion is systematically overestimated at low Galactic heights and slightly underestimated at high Galactic heights leading to an underestimate of the gradient of the dispersion with Galactic height. The causes of the systematics are revealed by repeating the calculation using a method inspired by \cite{Girard2006}. This method recovers the expected dispersion much more successfully and in particular yields a gradient of the dispersion with Galactic height which is approximately three times that found using the method presented by \citeauthor{MB2012}
\end{abstract}

\begin{keywords}
methods: data analysis - methods: numerical - Galaxy: kinematics and dynamics - Galaxy: solar neighbourhood
\end{keywords}

\section{Introduction}
\cite{MB2012b} used the Jeans equation to constrain the stellar mass density at heights $1.5\kpc\leq z\leq 4\kpc$ from the Galactic plane using a sample of 412 red giants and concluded that there is a lack of dark matter in the solar neighbourhood. \cite{BovyTremaine2012} have pointed out that one of the assumptions made by \cite{MB2012b} in their Jeans-equation analysis (specifically the assumption that the mean azimuthal velocity is independent of Galactocentric radius at all heights) is false and a reanalysis of the data without this assumption leads to a non-zero local dark matter density. Despite the fact that the data now appear to conform with standard expectation, it is wise to check that all parts of the analysis are sound before the matter is put to rest. To use the Jeans equation with confidence one needs to be sure that an underlying population has been isolated and that the mean velocities and velocity dispersions of that population can be reliably calculated. Here we investigate the method used by \citet[hereafter MB]{MB2012} to calculate the mean velocities and velocity dispersions of the thick disc. It is these quantities which are then used by \cite{MB2012b} and \cite{BovyTremaine2012} in their Jeans-equation analysis, so it is crucial that they are calculated correctly and that their associated errors are realistic.

MB use the probability plot method which we detail in Section~\ref{PPM}. We proceed by producing a sample drawn from the realistic Galaxy distribution functions of \cite{Binney2012b} chosen to be similar to the sample of MB. These distribution functions are briefly discussed in Section~\ref{BinneyModels}. In Section~\ref{MBMethod} we attempt to recover the known mean velocity and velocity dispersions of the thick disc using the method presented by MB. We also implement the error analysis used by MB, which does not include any estimates of the Poisson noise. In Section~\ref{PN} the analysis is repeated for a large number of samples to investigate the effects of the Poisson noise and to uncover any systematics introduced by the method. Finally in Section~\ref{Girard06}, the same data set is analysed by a similar method inspired by \cite{Girard2006} and the results are compared.

\section{Probability Plot Method}\label{PPM}
The data analysis used by MB uses the probability plot to determine the mean and standard deviation of a sample. Here we briefly present the method and give a simple example to demonstrate its use.

Suppose we have $N$ ordered data points. If these $N$ data points are drawn from a normal distribution of mean $\mu$ and standard deviation $\sigma$ then, for large $N$, the $i$th data point, $x_i$, should satisfy
\begin{equation}
F(x_i;\mu,\sigma) \simeq \frac{i}{N+1}.
\end{equation}
$F(x;\mu,\sigma)$ is the cumulative distribution function for a normal distribution, which is given by
\begin{equation}
F(x;\mu,\sigma) = \frac{1}{2}\Big[1+\mbox{erf}\Big(\frac{x-\mu}{\sqrt{2}\sigma}\Big)\Big],
\end{equation}
where $\mbox{erf}$ is the error function.
Hence the $i$th data point should lie at $c_i$ standard deviations from the mean where
\begin{equation}
c_i = \sqrt{2}\,\mbox{erf}^{-1}\Big(\frac{2i}{N+1}-1\Big)\approx \Big(\frac{x_i-\mu}{\sigma}\Big).
\label{Ci}
\end{equation}
Therefore a plot of $x_i$ against $c_i$ should have gradient $\sigma$ and intercept $\mu$, which may be found by linear regression. Such a plot is termed a probability plot.

\begin{figure}
$$\includegraphics{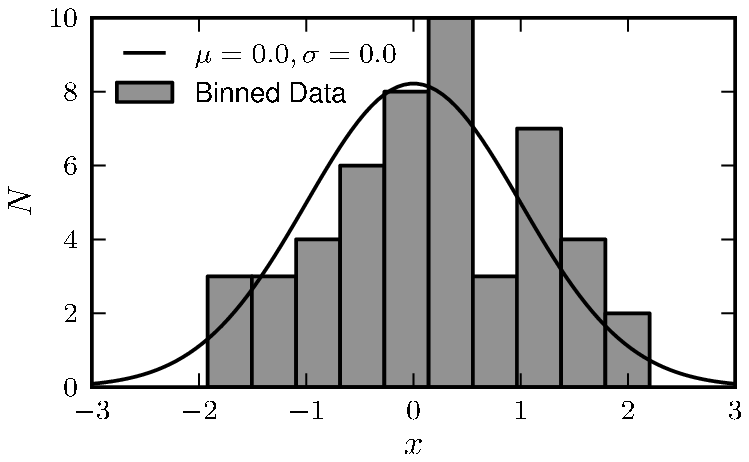}$$
$$\includegraphics{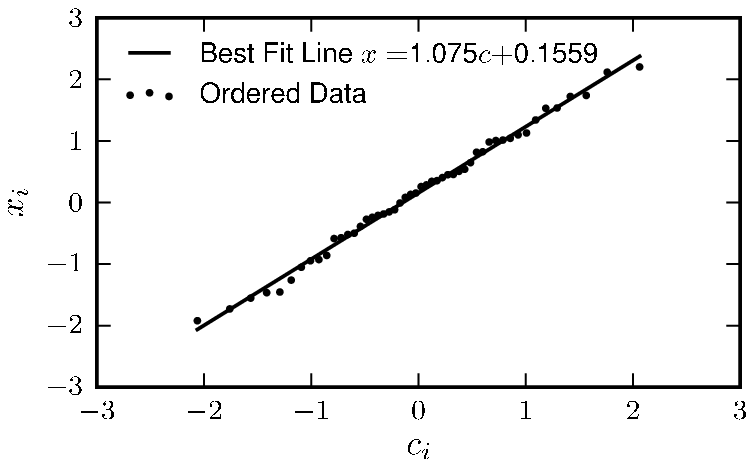}$$
\caption{Demonstration of finding the mean and standard deviation of a sample. The top panel shows a histogram of data drawn from a normal distribution of mean zero and unit standard deviation shown by the line. The bottom panel shows the ordered data plotted against the expected deviation in units of the standard deviation, $c_i$, given by Eq.~\ref{Ci}. A straight line fit to this plot yields an estimate for the standard deviation as the gradient and an estimate for the mean as the intercept. For this example the standard deviation is estimated as 1.075 and the mean as 0.156.}
\label{SimpleTest}
\end{figure}

Clearly this approach only works exactly if $N$ is large and the data have been drawn from a single normal distribution. However we can use it to estimate the mean and standard deviation of a sample drawn from any underlying distribution which is approximately Gaussian.
Here we show how the method operates for a simple case. We draw 50 data points from a normal distribution of mean $\mu=0$ and standard deviation $\sigma=1$, and assess how well the above method can recover these quantities. Fig.~\ref{SimpleTest} shows the result for one randomly drawn sample. The measured standard deviation for this sample is $s = 1.006$. From the linear regression the mean is estimated as $m = 0.156\pm0.141$ and the standard deviation $s = 1.075\pm0.152$ where the errors are given by the deviations of the points away from a straight line. With $N$ data points the expected error in the mean is $\sqrt{\sigma^2/N}$ and the expected error in the variance is approximately $\sqrt{2\sigma^4/N}$. Therefore, for this sample we expect an error in the mean of $0.141$ and an error in the standard deviation of $0.1$. This is well represented by the errors in the linear regression. Thus the method does work well when the data are drawn from an underlying Gaussian distribution and the errors from the linear regression are comparable to the expected Poisson noise.

When the data have been drawn from the sum of two Gaussian distributions, we can estimate the means and standard deviations by fitting straight line segments to different parts of the plot. This is the procedure followed by \cite{Bochanski2007} to calculate the velocity dispersions of the thin and thick discs using a tracer population of 7398 M dwarfs.

\section{Dynamical Galaxy models}\label{BinneyModels}
To test the method used by MB we require a sample drawn from a realistic Galaxy distribution function for which we know the underlying velocity moments. We use the models of \cite{Binney2012b}. These models were developed and discussed by \cite{Binney2010} and \cite{BinneyMcMillan2011}. The distribution function is a function of the actions in an axisymmetric potential: the radial action $J_r$, the vertical action $J_z$ and the $z$-component of the angular momentum, $L_z$. The advantage of this approach is that the distribution function clearly satisfies the Jeans' theorem as the actions are isolating integrals. These models consist of a thick and thin disc composed of quasi-isothermal distribution functions:
\begin{equation}
f(J_r,J_z,L_z) = f_{\sigma_r}(J_r,L_z)f_{\sigma_z}(J_z),
\end{equation}
where
\begin{equation}
f_{\sigma_r}(J_r,L_z) \equiv \frac{\Omega\Sigma}{\pi\sigma_r^2\kappa}\Bigg|_{R_c}[1+\tanh(L_z/L_0)]\mathrm{e}^{-\kappa J_r/\sigma_r^2}
\end{equation}
and
\begin{equation}
f_{\sigma_z}(J_z)\equiv \frac{\nu}{2\pi\sigma_z^2}\mathrm{e}^{-\nu J_z/\sigma_z^2}.
\end{equation}
Here $R_c(L_z)$ is the radius of a circular orbit with $z$-component of angular momentum, $L_z$. $\kappa(L_z)$, $\nu(L_z)$ and $\Omega(L_z)$ are the radial, vertical and circular epicycle frequencies respectively and $\Sigma(L_z)$ is the approximate surface density of the disc. The factor of $[1+\tanh(L_z/L_0)]$ eliminates retrograde stars. $\sigma_r(L_z)$ and $\sigma_z(L_z)$ are exponentially decaying function of $R_c$ which control the radial and vertical velocity dispersions:
\begin{equation}
\begin{split}
\sigma_r(L_z) &= \sigma_{r0}\mathrm{e}^{q(R_0-R_c)/R_d}\\
\sigma_z(L_z) &= \sigma_{z0}\mathrm{e}^{q(R_0-R_c)/R_d},
\end{split}
\end{equation}
where $\sigma_{r0}$ and $\sigma_{z0}$ are approximately equal to the radial and vertical velocity dispersions at the solar radius, $R_0$, and $R_d$ is the scale length of the disc.

\begin{figure}
$$\includegraphics{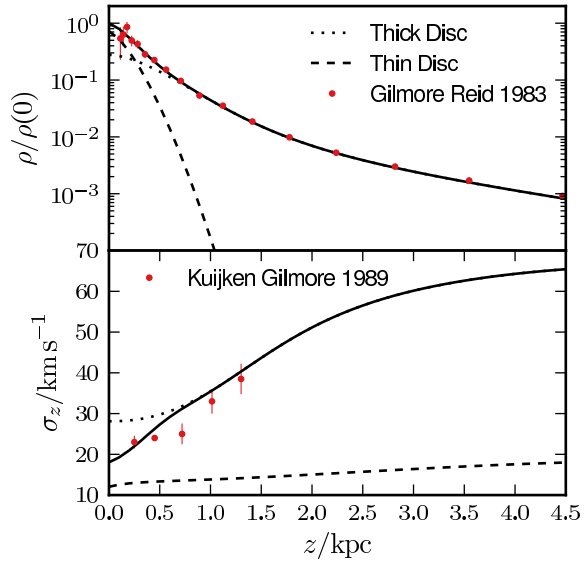}$$
\caption{Density profile and velocity dispersion for the Binney distribution function used in this paper as a function of Galactic height. The dashed line gives the contribution from the thin disc and the dotted line from the thick disc. The data points are taken from \citet{GilmoreReid1983} and \citet{KuijkenGilmore1989}.}
\label{DistFunc}
\end{figure}

The thick disc consists of a single quasi-isothermal distribution function, $f_{\rm thick}$, of fixed age whilst the thin disc is a superposition of quasi-isothermal distribution functions of differing ages from zero to $\tau_m=10\Gyr$. Following \cite{AumerBinney2009} we also adopt an age dependence for the velocity dispersion such that
\begin{equation}
\begin{split}
\sigma_r(L_z, \tau) &= \sigma_{r0}\Big(\frac{\tau+\tau_1}{\tau_m+\tau_1}\Big)^\beta\mathrm{e}^{q(R_0-R_c)/R_d}\\
\sigma_z(L_z, \tau) &= \sigma_{z0}\Big(\frac{\tau+\tau_1}{\tau_m+\tau_1}\Big)^\beta\mathrm{e}^{q(R_0-R_c)/R_d}.
\end{split}
\end{equation}
where we set $\beta=0.33$ and $\tau_1=0.11\Gyr$. We also assume a decreasing star formation rate with time with characteristic time $t_0=8\Gyr$ such that the full distribution for the thin disc is given by
\begin{equation}
f_{\rm thin} = \frac{\int_0^{\tau_m}\mathrm{d}\tau\,\mathrm{e}^{\tau/t_0}f_{\sigma_r}(J_R,L_z) f_{\sigma_z}(J_z)}{t_0(\mathrm{e}^{\tau_m/t_0}-1)}.
\end{equation}
The ratio of the thick to thin disc distribution functions is controlled by a parameter, $k_\mathrm{thk}=0.224$, which implies a fraction $0.224/1.224=0.18$ of the disc stars belong to the thick disc and implies a Solar neighbourhood ratio of 0.28. We set the parameters to the values given in Table~\ref{ParamTable}. The density profile and velocity dispersion for this distribution function as a function of Galactic height are shown in Fig.~\ref{DistFunc}.

\begin{table}
\caption{Parameters for the Binney distribution function used throughout the paper. Velocity dispersions have units $\kms$ and scale lengths have units $\kpc$.}
\label{ParamTable}
\begin{center}
\begin{tabular}{lll}
\hline
Thin&$\sigma_{r0}$&42.3\\
&$\sigma_{z0}$&20.3\\
&$R_d$&2.17\\
&$q$&.040\\
\hline
Thick &$\sigma_{r0}$&26.3\\
&$\sigma_{z0}$&34.0\\
&$R_d$&3.66\\
&$q$&1.068\\
&$k_{\rm thk}$&0.224\\
\hline
\end{tabular}
\end{center}
\end{table}

We limit our investigation to just the $W$ component of the velocity (the component along the $z$ direction). We are not interested in the full distribution function so we marginalise over the other two velocity components ($U$ and $V$) to find the number of stars per unit $W$ velocity per unit volume as
\begin{equation}
n_W(R,z,W) = \int_{-\infty}^\infty \mathrm{d}U \int_{-\infty}^\infty \mathrm{d}V f(J_r,J_z,L_z)
\end{equation}
where in practice the limits of the integrals are finite as the distribution function falls off rapidly at large velocities. The transformation from polar positions and velocities to actions is carried out by the algorithm presented by \cite{Binney2012a}.  We use an adjusted version of Potential II from \cite{DB1998} which consists of a thin and thick disc, a gas disc and two spheroids representing the bulge and the halo. We have increased the scale-height of the thin disc to $360\pc$ and increased the mass of the thin disc such that the circular velocity at the solar radius is $220\kms$.

From this distribution function we are able to draw a sample of stars. The sample is selected by following a rejection algorithm. If we wish to draw stars which all lie at the solar radius, $R_0$, and which lie in the ranges $z_{\rm min}\leq z \leq z_{\rm max}$ and $|W|\leq W_{\rm max}$ then we first note that the maximum of the distribution function in this range occurs at $z=z_{\rm min}$ and $W=0$. This gives us a normalisation. We then proceed by drawing trial values of $z=z_t$ and $W=W_t$ from uniform distributions over the required ranges and accepting this trial as a data point with probability $n_W(R_0,z_t,W_t)/n_W(R_0,z_{\rm min},0)$.

The obvious advantage of drawing sample data from a known distribution function is that we know exactly the underlying properties of the distribution. In this paper we focus on calculating the mean $W$ velocity and the $W$ velocity dispersion of the thick disc, $\sigma_z$. As the distribution function is a symmetric function of $W$ we expect $\langle W\rangle=0$. $\sigma_z$ is given by
\begin{equation}
\sigma_z^2(R,z) =  \frac{\int_{-\infty}^\infty \mathrm{d}U \int_{-\infty}^\infty \mathrm{d}V  \int_{-\infty}^\infty \mathrm{d}W\,W^2f_{\rm thick}(J_r,J_z,L_z)}
{\int_{-\infty}^\infty \mathrm{d}U \int_{-\infty}^\infty \mathrm{d}V  \int_{-\infty}^\infty \mathrm{d}W\,f_{\rm thick}(J_r,J_z,L_z)}.
\end{equation}

\section{MB method}\label{MBMethod}
\begin{figure}
$$\includegraphics{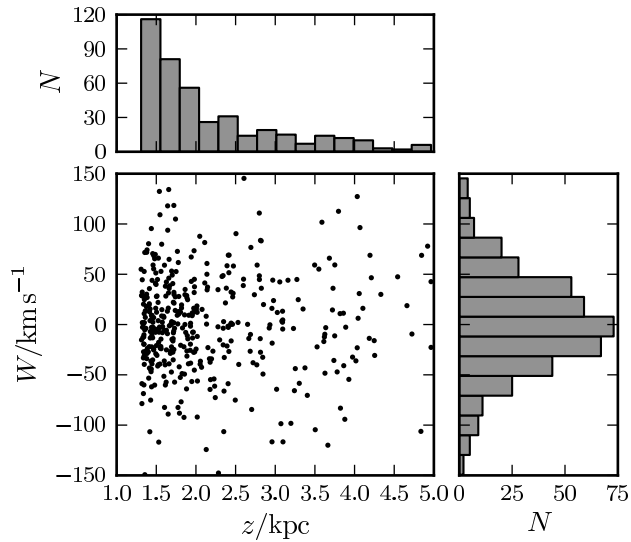}$$
\caption{Sample of 412 stars drawn from the Binney distribution function that emulates the MB sample.}
\label{SampleHists}
\end{figure}

MB use a sample of 412 red giants which lie in the range $1.3\kpc \leq z \leq 5\kpc$ and $|W|\leq150\kms$. The cut in $W$ is performed to remove contamination from the halo. MB estimate an error of $0.7\kms$ in the radial velocity measurements and an error of approximately $20\percent$ in the distances\footnote{MB state that there is an additional $10-20\percent$ systematic error in the distances to thin disc stars due to the thin disc stars not following the assumed age and metallicity distributions. We ignore this additional error here. This error will increase thin disc contamination at low $z$ but should not affect the determination of the velocity dispersion or conclusions presented here significantly.}. We use the procedure outlined above to draw 412 stars from the Binney distribution function which all lie at the solar radius and inside the range probed by MB. As the $W$ velocity is nearly entirely radial velocity error we include a random Gaussian error of $0.7\kms$ to the $W$ velocities and we assume that the full distance error of $20\percent$ corresponds to a $20\percent$ error in the $z$ values. In order to correctly account for stars which may have entered our sample due to the error in their distances we increase the sample range to $1\kpc \leq z \leq 6\kpc$ and then cut out any stars which lie outside $1.3\kpc \leq z \leq 5\kpc$ after the error has been included. Histograms of the resulting sample are shown in Fig.~\ref{SampleHists}.

We now follow the same procedure as MB to extract the mean $W$ velocity and the $W$ velocity dispersion, $\sigma_z$. We first bin the data in $z$ with bin centres spaced by $0.1\kpc$ in the range $1.5\kpc\leq z \leq4.5\kpc$. The bin sizes are allowed to vary such that we have 100 data per bin for $z\leq2.1\kpc$, 80 data per bin for $2.2\kpc\leq z\leq 2.4\kpc$ and 50 data per bin for $z\geq2.5\kpc$. For each binned subset of the sample we follow the method outlined in Section~\ref{PPM} by ordering the sample in $W$ and performing a linear regression between the sample velocities and the expected deviations to find the mean velocity of the bin and the velocity dispersion.

For $z\leq2.5\kpc$ we expect a non-negligible thin disc contamination. MB try to isolate the thick disc contribution by only fitting the wings of the distribution where, as the thick disc has a higher velocity dispersion than the thin disc, the data are assumed to be contributed by thick disc stars. Therefore, we first sort all the data in the bin and assign each an expected deviation but only fit a straight line to the data which have $|W|>30\kms$. Each wing is fitted separately. MB do not make it completely clear how they combine the fits of each wing but here we adopt the procedure of fitting each wing independently and then calculating the mean and standard deviation by an average of the intercepts and gradients respectively.

MB also ignore any points which seem to be outliers in the probability plots when performing the linear regression. We simulate this effect by ignoring the most negative and most positive data point when fitting a straight line to the probability plot.

\subsection{Errors}\label{Errors}
\subsubsection{Observational Errors}
\begin{figure}
$$\includegraphics{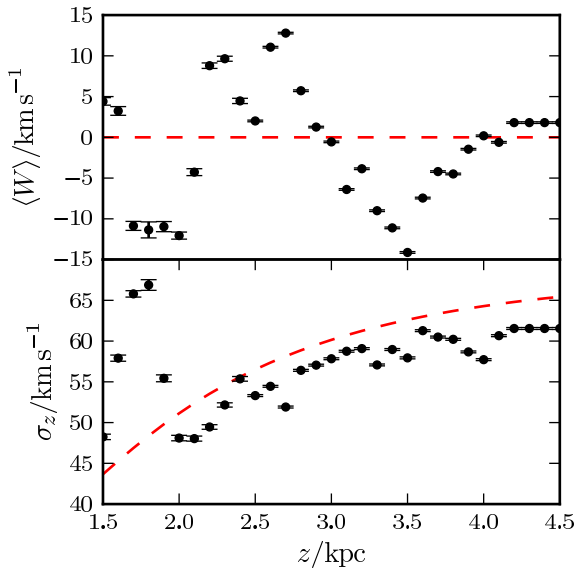}$$
\caption{Mean $W$ velocity and $W$ velocity dispersion, $\sigma_z$, against height above Galactic plane. The points give the values calculated in each bin using the MB method along with the errors estimated by 1000 samples adding random errors of $0.7\kms$ to the $W$ velocity. The red dashed lines show the exact mean velocity and velocity dispersion calculated from the Binney distribution function from which the sample was drawn. Clearly the error bars do not give a good estimate of the deviation from the true value.}
\label{SmallErrors}
\end{figure}

MB estimate the errors in their calculated $W$ moments by essentially only considering the observational error in the $W$ velocity as follows. In each bin MB add random Gaussian errors for the distance and radial velocity to each data point to generate 1000 samples but do not re-bin the data at all. The errors are then estimated as the standard deviation of the estimates obtained from each sample. Following MB we take the $W$ velocity error to be $0.7\kms$. The results of this procedure are shown in Fig.~\ref{SmallErrors}. The error bars are very small giving the impression we have very precise results. However the data are clearly scattered around the true result by amounts much greater than the error bars. This is because we have ignored two much larger sources of error: the errors in the distances moving stars from bin to bin and the Poisson noise.

\begin{figure}
$$\includegraphics{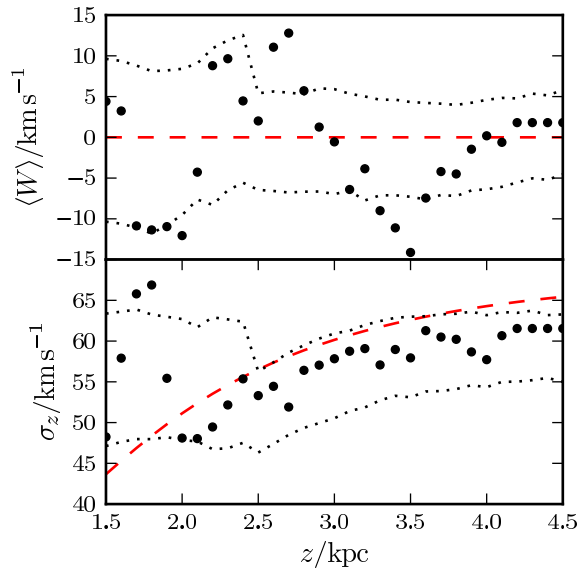}$$
\caption{Mean $W$ velocity and $W$ velocity dispersion, $\sigma_z$, against height above Galactic plane for the sample of 412 stars. The points give the values calculated in each bin using the MB method. The dotted lines show the one standard deviation limits. These were estimated by calculating the standard deviations of the calculated values for 1000 samples formed by adding random errors of $0.7\kms$ to the $W$ velocity and a $20\percent$ error on the distance with re-binning. The red dashed lines show the exact mean velocity and velocity dispersion calculated from the Binney distribution function from which the sample was drawn.}
\label{BiggerErrors}
\end{figure}

MB estimate the distance error to be approximately $20\percent$. As well as a $W$ velocity error, we add a random Gaussian error of $20\percent$ to the $z$ coordinates of the data, re-bin the data and recalculate the mean velocities and dispersions. Repeating this 1000 times we calculate the errors as the standard deviations of the estimates. These results are shown in Fig.~\ref{BiggerErrors}. The observational errors are now much larger and the results are consistent with the truth within the errors. We have not yet made any estimate for the Poisson noise of the estimate but it seems that, as the data are consistent with the truth, the observational errors are of the same order as the Poisson noise.

One standard deviation lines are shown in Fig.~\ref{BiggerErrors}. Adding errors in $z$ to the data shifts stars from bin to bin. With such a large distance error we have many samples in which higher velocity stars have been pulled down to lower Galactic heights and lower velocity stars are displaced to greater heights. This has the effect of flattening the velocity distribution when averaged over the many samples.

\subsubsection{Poisson Noise and systematics}\label{PN}
\begin{figure}
$$\includegraphics{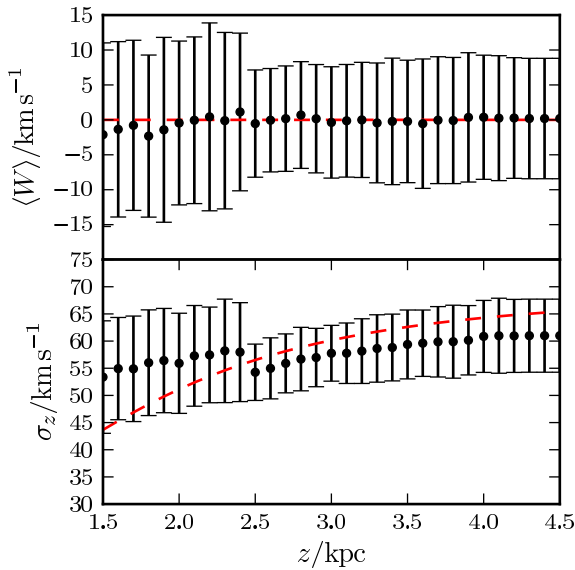}$$
\caption{MB method: Average mean $W$ velocity and $W$ velocity dispersion, $\sigma_z$, against height above Galactic plane for the 100 samples of 412 stars. The points give the average values calculated in each bin using the MB method. The error bars show the standard deviation of the calculated values in each bin. The red dashed lines show the velocity moments of the thick disc calculated using the underlying  distribution function. The dispersion is overestimated at low $z$ and slightly underestimated at high $z$ leading to an underestimated gradient.}
\label{PoissonNoise}
\end{figure}

With such a small sample it is difficult to disentangle the Poisson noise from systematic errors arising from the MB procedure. However, as we have direct access to the distribution function we can estimate the Poisson noise by repeatedly drawing samples from the distribution function and evaluating the observables for each sample. Therefore we draw 100 samples of 412 stars and repeat the above procedure for each sample. We then estimate the average sample mean and dispersion in each bin along with the Poisson error in both quantities by calculating the standard deviations. The results are shown in Fig.~\ref{PoissonNoise}. 

The calculation of the mean $W$ velocity is entirely consistent with being zero as required, but the recovery of the $W$ velocity dispersion curve is less successful. For low Galactic heights we are overestimating the velocity dispersion whilst for larger Galactic height we are slightly underestimating the dispersion.

For low $z$ we are ignoring all data for which $|W|<30\kms$ when fitting a straight line to the data. This means we give more weight to data with higher $W$ velocities and so the distribution seems broader than it actually is. The probability plot is particularly sensitive at the wings. If we consider an ordered data set that is drawn from a known underlying Gaussian distribution, we can assign each a value of $c_i$ by the method outlined in Section~\ref{PPM}. If we add a single point which is smaller than all the other data points but still drawn from the underlying distribution the probability of the new point lying above the line with correct mean and standard deviation, but still lower than its neighbouring data point, is equal to the probability of it lying beneath the line. As there is a much larger range of values below the line than above, the estimated gradient in this region will in general be overestimated. We need to use a sufficient number of stars to perform the linear fit in order to reduce this effect.

At high $z$ we have very few stars in the sample so in order to fill the bin with enough stars we must include stars at lower $z$. In general these stars have smaller velocities and so the resulting velocity dispersion for the bin is reduced. A very minor effect may also be due to removing stars which have $|W|>150\kms$ to avoid halo contaminants, so the distribution is unfairly weighted by low-velocity stars and the dispersion is underestimated. From Fig.~\ref{PoissonNoise} we can perform a simple linear fit to the $(z,\sigma_z)$ plot to find that the data points imply a gradient of ${\mathrm{d}\sigma_z}/{\mathrm{d}z=2.25\kms\kpc^{-1}}$. A linear fit to the true dispersion curve gives a gradient of ${\mathrm{d}\sigma_z}/{\mathrm{d}z=6.82\kms\kpc^{-1}}$ so the MB method underestimates the gradient by a factor of three.

\begin{figure}
$$\includegraphics{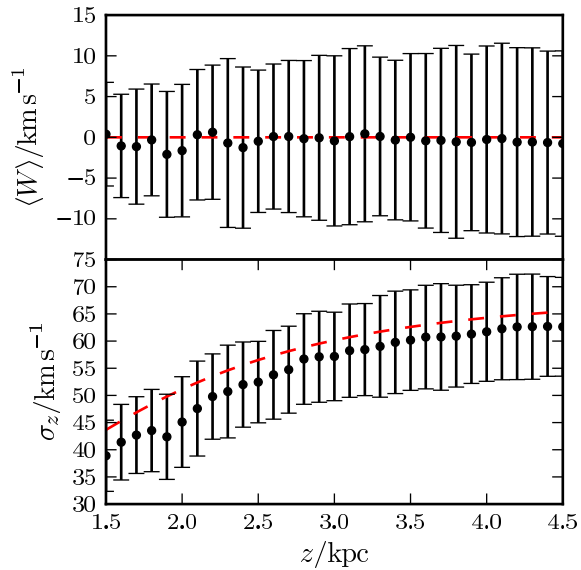}$$
\caption{\citeauthor{Girard2006} method: average mean $W$ velocity and $W$ velocity dispersion, $\sigma_z$, against height above Galactic plane for the 100 samples of 412 stars. The points give the average values calculated in each bin using a method inspired by \citet{Girard2006} outlined in Section~\ref{Girard06}. The error bars show the standard deviation of the calculated values in each bin. The red dashed lines show the velocity moments of the thick disc calculated using the underlying  distribution function. The dispersion is slightly underestimated due to the neglected thin disc contamination and preferentially sampling stars from lower Galactic heights, but in general the recovery of the truth is more successful than using the MB method. In particular the calculated gradient in $\sigma_z$ is approximately a factor of three larger.
  }
\label{Girard}
\end{figure}

\section{Comparison with other work}\label{Girard06}
MB state that their gradient of $\sigma_z$ with $z$ is shallower than previous authors' work. The sample studied by MB is a subset of the sample studied by \citet[hereafter G06]{Girard2006}. These authors found a gradient of the $V$ velocity dispersion a factor of two higher than MB, which MB claim is due to G06 not removing thin disc and halo contaminants in their analysis. We conclude by following a method inspired by the method used by G06 to test whether their results are more secure than those reported by MB. The G06 method is very similar to that used by MB. The authors have a sample of approximately 1200 stars. They split each data point into 100 subunits to account for the distance error and form bins of 100 subunits to estimate the $U$ and $V$ velocity dispersion. The probability plot method is used to estimate the dispersion but crucially only the central $80\percent$ of the data is used in the linear fit and the central region is not excluded for any of the bins. We follow a similar, but simpler, method on each of the 100 samples of 412 stars. We do not split the data points into subunits. We use bins spaced by $0.1\kpc$ containing 32 data points, as we have only a third of the number of data points in the G06 sample, and we use only the central $80\percent$ of the data in each bin for the linear fit in the probability plot method.  Fig.~\ref{Girard} shows the result of this experiment. 

The estimate of the velocity dispersion provided by the G06 method is more reliable than the MB method. At low $z$ the dispersion is now marginally underestimated which is to be expected due to the thin disc contamination. However, even with fewer stars in the bin, the error in the dispersion at low $z$ is smaller than the equivalent errors in the MB method. This is a clear reflection of the dangers of only using the wings of the distribution to calculate the dispersion. At high $z$ the problem of preferentially sampling stars at lower heights seems to also have been slightly reduced as the bin size is small enough for the dispersion to be calculated using only local stars. A simple linear fit to the $(z,\sigma_z)$ plot gives a gradient of ${\mathrm{d}\sigma_z}/{\mathrm{d}z}=7.87\kms\kpc^{-1}$.

\section{Conclusions}
We have drawn a sample of 412 stars from the distribution function of \cite{Binney2012b}. The sample was chosen to replicate the sample presented by MB. We performed the same procedure as MB to extract the mean vertical velocity and vertical velocity dispersion of the thick disc as a function of Galactic height and compared it to the known moments of the vertical velocity of the thick disc of the underlying distribution. We find that the variation of the dispersion with $z$ is far noisier than that found by MB. The majority of the error arises from the Poisson noise inherent in the limited sample size. We show that the observational errors in the velocities, which MB claim as the total error, cannot account for the deviation. A larger source of error arises from distance errors moving stars from bin to bin and we show that this is of a similar order to the Poisson noise.

A large number of samples reveals that the method systematically overestimates the dispersion at low $z$ and underestimates it at high $z$. The two effects combined lead to a flatter curve of dispersion against $z$. We recalculated the mean velocity and velocity dispersion using a method inspired by G06 which reveals the causes of these two effects: at low $z$, only fitting the wings of the distribution to remove thin disc contaminants makes the distribution appear broader, and at high $z$, large bin sizes preferentially sample stars at lower Galactic height which in general have a lower velocity. The G06 method produces a much better fit to the expected velocity dispersion and there is approximately a factor of three discrepancy in the gradient of the dispersion as a function of $z$ between the results of the MB and G06 methods. This discrepancy is not a result of more precise measurements or analysis but purely a result of systematics in the data analysis introduced by MB. We have not touched upon the values given by MB for the $U$ and $V$ dispersions but similar effects are expected to occur.

The results presented here should serve as a useful demonstration of the expected errors and potential biases which arise when using a method similar to the MB method. We have demonstrated the need to understand the errors and systematics of methods which are applied to observational data and that pseudo-samples from realistic Galaxy distribution functions are a useful tool in this respect. The effect that the biases and errors demonstrated in this paper have on the dark matter mass estimates \citep{MB2012b,BovyTremaine2012} is beyond the scope of this paper. However, the results of this paper are relevant to both these mass estimate determinations and more generally to the understanding of Galactic disc kinematics.

\section*{Acknowledgements}
I thank James Binney for carefully reading a draft of this work and providing his distribution function code. I also thank the members of the Oxford dynamics group for valuable conversations and acknowledge the support of the STFC.

\bibliographystyle{mn2e-2}
\bibliography{VelocityDispersionDetermination-JasonSanders.bib}

\begin{thebibliography}{13}
\expandafter\ifx\csname natexlab\endcsname\relax\def\natexlab#1{#1}\fi

\bibitem[{{Aumer} \& {Binney}(2009)}]{AumerBinney2009}
{Aumer} M., {Binney} J.~J., 2009, \mnras, 397, 1286

\bibitem[{{Binney}(2010)}]{Binney2010}
{Binney} J., 2010, \mnras, 401, 2318

\bibitem[{{Binney}(2012{\natexlab{a}})}]{Binney2012b}
{Binney} J., 2012{\natexlab{a}}, \mnras, submitted

\bibitem[{{Binney}(2012{\natexlab{b}})}]{Binney2012a}
{Binney} J., 2012{\natexlab{b}}, \mnras, submitted

\bibitem[{{Binney} \& {McMillan}(2011)}]{BinneyMcMillan2011}
{Binney} J., {McMillan} P., 2011, \mnras, 413, 1889

\bibitem[{{Bochanski} {et~al}\mbox{.}(2007){Bochanski}, {Munn}, {Hawley},
  {West}, {Covey}, \& {Schneider}}]{Bochanski2007}
{Bochanski} J.~J., {Munn} J.~A., {Hawley} S.~L., {West} A.~A., {Covey} K.~R.,
  {Schneider} D.~P., 2007, \aj, 134, 2418

\bibitem[{{Bovy} \& {Tremaine}(2012)}]{BovyTremaine2012}
{Bovy} J., {Tremaine} S., 2012, \apj, submitted

\bibitem[{{Dehnen} \& {Binney}(1998)}]{DB1998}
{Dehnen} W., {Binney} J., 1998, \mnras, 294, 429

\bibitem[{{Gilmore} \& {Reid}(1983)}]{GilmoreReid1983}
{Gilmore} G., {Reid} N., 1983, \mnras, 202, 1025

\bibitem[{{Girard} {et~al}\mbox{.}(2006){Girard}, {Korchagin},
  {Casetti-Dinescu}, {van Altena}, {L{\'o}pez}, \& {Monet}}]{Girard2006}
{Girard} T.~M., {Korchagin} V.~I., {Casetti-Dinescu} D.~I., {van Altena} W.~F.,
  {L{\'o}pez} C.~E., {Monet} D.~G., 2006, \aj, 132, 1768

\bibitem[{{Kuijken} \& {Gilmore}(1989)}]{KuijkenGilmore1989}
{Kuijken} K., {Gilmore} G., 1989, \mnras, 239, 605

\bibitem[{{Moni Bidin} {et~al}\mbox{.}(2012{\natexlab{a}}){Moni Bidin},
  {Carraro}, \& {M{\'e}ndez}}]{MB2012}
{Moni Bidin} C., {Carraro} G., {M{\'e}ndez} R.~A., 2012{\natexlab{a}}, \apj,
  747, 101

\bibitem[{{Moni Bidin} {et~al}\mbox{.}(2012{\natexlab{b}}){Moni Bidin},
  {Carraro}, {M{\'e}ndez}, \& {Smith}}]{MB2012b}
{Moni Bidin} C., {Carraro} G., {M{\'e}ndez} R.~A., {Smith} R.,
  2012{\natexlab{b}}, \apj, 751, 30

\end{thebibliography}

\bsp

\label{lastpage}

\end{document}